\documentclass[12pt,preprint]{aastex}

\usepackage{amsmath}
\usepackage{bm}
\slugcomment{\today}

\shorttitle{New equation of state with nuclei and hyperons}
\shortauthors{Miyatsu et al.}

\begin{document}


\title{A new equation of state for neutron star matter with nuclei \\ in the crust and hyperons in the core}


\author{Tsuyoshi Miyatsu\altaffilmark{1}, Sachiko Yamamuro\altaffilmark{2}, and Ken'ichiro Nakazato\altaffilmark{2}}

\email{tmiyatsu@ssu.ac.kr}


\altaffiltext{1}{Department of Physics, Soongsil University, Seoul 156-743, Korea}
\altaffiltext{2}{Department of Physics, Faculty of Science \& Technology, Tokyo University of Science, 2641 Yamazaki, Noda, Chiba 278-8510, Japan}


\begin{abstract}
% We will write abstract here.
The equation of state for neutron stars in a wide-density range at zero temperature is constructed. The chiral quark-meson coupling model within relativistic Hartree-Fock approximation is adopted for uniform nuclear matter. The coupling constants are determined so as to reproduce the experimental data of atomic nuclei and hypernuclei. In the crust region, nuclei are taken into account within the Thomas-Fermi calculation. All octet baryons are considered in the core region, while only $\Xi^{-}$ appears in neutron stars. The resultant maximum mass of neutron stars is $1.95M_\odot$, which is consistent with the constraint from the recently observed massive pulsar, PSR J1614-2230.
\end{abstract}


%% subject header.  Objects should be in the appropriate "individual"
%% headers (e.g. quasars: individual, stars: individual, etc.) with the
%% additional provision that the total number of headers, including each
%% individual object, not exceed six.  The \objectname{} macro, and its
%% alias \object{}, is used to mark each object.  The macro takes the object
%% name as its primary argument.  This name will appear in the paper
%% and serve as the link's anchor in the electronic edition if the name
%% is recognized by the data centers.  The macro also takes an optional
%% argument in parentheses in cases where the data center identification
%% differs from what is to be printed in the paper.

\keywords{dense matter --- equation of state --- stars: neutron}


\section{Introduction}\label{intro}

Understanding of the equation of state (EOS) for dense matter is important to clarify compact astrophysical phenomena. For instance, neutron stars are composed of the core with supra-nuclear densities and crust with subnuclear densities \citep[e.g.,][]{Glendenning:1997wn,Lattimer:2004pg,Lattimer:2006xb}. The mass and radius of neutron stars are mainly determined by the EOS at core. Nevertheless, the crust is also important to account for some phenomena such as the pulsar glitches \citep[][]{AI75,Link99,Nils12} or the quasi-periodic oscillations in giant flares emitted by highly-magnetized neutron stars \citep[][]{stroh06,Sotani:2012qc}. Therefore, the range of densities required to investigate neutron stars is enormous. Furthermore, to study supernova explosions, protoneutron stars (nascent neutron stars from supernovae), or binary neutron star mergers, we need an EOS with not only a wide-density range but also finite temperatures.

At present, there are a few types of nuclear EOSs which has been widely used in astrophysical simulations. One of the most popular EOSs is calculated with the relativistic mean-field (RMF) theory for uniform nuclear matter and the Thomas-Fermi model for nonuniform matter \citep[][]{Shen:1998gq,Shen:1998by}. The EOS of \citet{Lattimer:1991nc}, which is based on Skyrme-type nuclear interactions and the compressible liquid-drop model, has also been used over the past decade. Moreover, recently, several EOSs based on the RMF model have been opened to the public \citep[e.g.,][]{Hempel:2009mc,Shen:2010pu}. Incidentally, there is an attempt to establish the EOS with the cluster variational method for nuclear matter \citep[][]{kann07,Kanzawa:2009uc}. Unfortunately, however, these EOSs do not include hyperons. According to terrestrial nuclear experiments such as heavy-ion collisions in J-PARC and GSI-FAIR, there are many evidences of existing hyperons \citep[e.g.,][]{nagae10,botta12}. Roughly speaking, since the mass differences between hyperons and nucleons are comparable to the nucleon Fermi energy in neutron stars, hyperons should be considered in an EOS for neutron stars.

These days, thanks to advances in observations, it would become possible astrophysically to get some information on the properties of nuclear matter. In particular, the discovery of a so-called two-solar-mass neutron star (PSR J1614-2230, $1.97\pm0.04M_{\odot}$) puts a severe constraint on the EOS for nuclear matter \citep[][]{Demorest:2010bx}. Generally speaking, the inclusion of other degrees of freedom, such as hyperon admixture, meson condensation and quark matter, softens an EOS, and the maximum mass of neutron stars is thus reduced \citep[e.g.,][]{Glendenning:1991es,glend98,weber05,SchaffnerBielich:2008kb}. In fact, the mass of J1614-2230 cannot be accounted for extending the EOS of \citet{Shen:1998gq,Shen:1998by} to a hyperon inclusion \citep[][]{Ishizuka:2008gr,Shen:2011qu} or a hadron-quark phase transition \citep[][]{self08a}. Thus, it is urgent to construct an EOS which is consistent with both of the terrestrial nuclear experiments and the astrophysical observations.

To solve the ``hyperon puzzle'', one of us (T.M.) introduced the framework based on the Hartree-Fock calculation in RMF theory \citep[][]{Miyatsu:2011bc,Katayama:2012ge}. An EOS becomes stiff due to the exchange (Fock) term, which is not taken into account in the previous studies \citep[][]{Ishizuka:2008gr,Shen:2011qu}. They showed that it is possible to make an EOS with hyperons which is consistent with the mass of J1614-2230, using the hyperon interaction based on the experimental hyperon potentials \citep[][]{sb96,Ishizuka:2008gr,botta12} and hyperon-nucleon scattering data \citep[][]{Rijken:2010zzb}. Among the models proposed in \citet{Katayama:2012ge}, they recommended the chiral quark-meson coupling (CQMC) model which was shown as case (d) in Table~8 of their paper. In this model, the internal structure of baryons is considered based on the quark degrees of freedom and the quark-quark hyperfine interactions, and its modification in the nuclear medium is effectively taken into account as a density dependence of an interaction. Hereafter, we denote this model as KMS12.

We extend the EOS of KMS12 for use in astrophysics. As a first step, in this paper, we fix the model which is consistent with the experimental data on atomic nuclei. Here we readjust the coupling constants in KMS12 so as to reproduce the gross feature of nuclear mass data with the Thomas-Fermi calculation, as done in \citet{Kanzawa:2009uc}. Lately, finite nuclei are dealt within the framework of RMF model. Nevertheless, the final goal of us is an EOS for a wide range of parameters including finite temperatures. The Thomas-Fermi approximation is applicable for this purpose \citep[][]{Shen:1998by}, and we adopt it also for the calibration of the coupling constants self-consistently. Moreover, in this paper, we construct the EOS for low-density nonuniform matter at zero temperature and apply the result to neutron stars including hyperons. Consequently we can account for a neutron star with $1.95M_\odot$, which lies in the $1\sigma$ limit on the mass of J1614-2230 \citep[][]{Demorest:2010bx}.

This paper is arranged as follows. In Section~\ref{eos}, a brief review of the formalism for the Hartree-Fock calculation in RMF theory with the CQMC model is presented. We show the method to determine the coupling constants from the experimental data of atomic nuclei and hypernuclei in Section~\ref{cplcnst}. Application for neutron stars and discussions are addressed in Section~\ref{nstar}. Finally, summary is included in Section~\ref{summary}. In this paper, the velocity of light $c$ and the Planck constant $\hbar$ are taken to be unity.

\section{Hartree-Fock calculation in relativistic mean-field theory}\label{eos}

In this section, we summarize the formulations of our EOS for uniform nuclear matter \citep[][]{Miyatsu:2011bc,Katayama:2012ge}. Here we adopt the RMF theory, in which baryons interact via the exchange of mesons. In order to calculate the interactions, we take into account not only the tadpole (Hartree) but also exchange (Fock) diagrams. We call this method relativistic Hartee-Fock (RHF) approximation. Furthermore we consider the baryon structure variation due to the interaction in matter using the CQMC model. In this model, the properties of nuclear matter can be self-consistently calculated by the coupling of scalar and vector fields to the quarks within nucleons \citep[][]{Guichon:1987jp,Saito:1994ki} and the quark-quark hyperfine structures due to the exchanges of gluon and pion are included based on chiral symmetry \citep[][]{Guichon:2008zz,Nagai:2008ai,Miyatsu:2010zz}. 

We start the descriptions of our EOS with the Lagrangian density for hadronic matter:
\begin{equation}
	\mathcal{L}_{H} = \mathcal{L}_{B} + \mathcal{L}_{M} + \mathcal{L}_{int}.
	\label{eq:total-Lagrangian}
\end{equation}
The first term is the baryon term and given by
\begin{equation}
	\mathcal{L}_{B} = \sum_{B} \bar{\psi}_{B} \left( i\gamma_{\mu}\partial^{\mu}-M_{B} \right) \psi_{B},
	\label{eq:Lagrangian-kinetic}
\end{equation}
where $\psi_{B}$ is the baryon field and $M_{B}$ is the baryon mass in vacuum. The sum $B$ runs over the octet baryons, $p$, $n$, $\Lambda$, $\Sigma^{+0-}$ and $\Xi^{0-}$. Proton and neutron are denoted collectively as $N$. For the free baryon masses, we take $M_{N}=939$~MeV, $M_{\Lambda}=1116$~MeV, $M_{\Sigma}=1193$~MeV and $M_{\Xi}=1313$~MeV \citep[][]{Pal:1999sq}. Taking into account the isoscalar ($\sigma$ and $\omega$) mesons and the isovector (${\bm \pi}$ and ${\bm \rho\,}$) mesons, we write the meson term as
\begin{eqnarray}
	\mathcal{L}_{M}
	&=& \frac{1}{2}\left(\partial_{\mu}\sigma\partial^{\mu}\sigma-m_{\sigma}^{2}\sigma^{2}\right)
	+ \frac{1}{2}m_{\omega}^{2}\omega_{\mu}\omega^{\mu} 
	- \frac{1}{4}W_{\mu\nu}W^{\mu\nu} \nonumber \\
	&+& \frac{1}{2}m_{\rho}^{2}\bm{\rho}_{\mu}\cdot\bm{\rho}^{\,\mu}
	- \frac{1}{4}\bm{R}_{\mu\nu}\cdot\bm{R}^{\mu\nu}
	+ \frac{1}{2}\left(\partial_{\mu}\bm{\pi}\cdot\partial^{\mu}\bm{\pi}-m_{\pi}^{2}\bm{\pi}^{2}\right),
	\label{eq:Lagrangian-meson}
\end{eqnarray}
with
\begin{eqnarray}
	W_{\mu\nu} &=& \partial_{\mu}\omega_{\nu} - \partial_{\nu}\omega_{\mu}, 
	\label{eq:covariant-derivative-omega} \\
	\bm{R}_{\mu\nu} &=& \partial_{\mu}\bm{\rho}_{\nu} - \partial_{\nu}\bm{\rho}_{\mu},
	\label{eq:covariant-derivative-rho}
\end{eqnarray}
where the meson masses are chosen as $m_{\sigma}=550$~MeV, $m_{\omega}=783$~MeV, $m_{\pi}=138$~MeV and $m_{\rho}=770$~MeV \citep[][]{Pal:1999sq}.

The interaction Lagrangian is given by
\begin{eqnarray}
	\mathcal{L}_{int} &=& \sum_{B}\bar{\psi}_{B}\left[g_{\sigma B}(\sigma) \sigma
	-g_{\omega B}\gamma_{\mu}\omega^{\mu} 
	+ \frac{f_{\omega B}}{2\mathcal{M}}\sigma_{\mu\nu}\partial^{\nu}\omega^{\mu}\right. \nonumber \\
	&-& \left. g_{\rho B}\gamma_{\mu}\bm{\rho}^{\,\mu}\cdot\bm{I}_B
	+ \frac{f_{\rho B}}{2\mathcal{M}}\sigma_{\mu\nu}\partial^{\nu}\bm{\rho}^{\,\mu}\cdot\bm{I}_B
	- \frac{f_{\pi B}}{m_{\pi}}\gamma_{5}\gamma_{\mu}\partial^{\mu}\bm{\pi}\cdot\bm{I}_B \right]\psi_{B},
	\label{eq:Lagrangian-interaction}
\end{eqnarray}
where the common scale mass $\mathcal{M}$ is taken to be the free nucleon mass \citep[][]{Rijken:2010zzb} and the commutation operator for $\gamma$ matrix is given by $\sigma_{\mu\nu}=i\left[\gamma_{\mu},\gamma_{\nu}\right]/2$. We denote the isospin matrix for baryon $B$ as $\bm{I}_B$, which is set to $\bm{I}_B=0$ for iso-singlet baryons. While $f_{\omega B}$ and $f_{\rho B}$ are the isoscalar-tensor and isovector-tensor coupling constants, the terms with $g_{\sigma B}(\sigma)$, $g_{\omega B}$, $g_{\rho B}$ and $f_{\pi B}$ correspond to $\sigma$-, $\omega$-, $\rho$-, $\pi$-$B$ couplings, respectively. We settle $g_{\omega B}$, $g_{\rho B}$, $f_{\omega B}$, $f_{\rho B}$ and $f_{\pi B}$ to be constants. Nevertheless, it is not the case for $g_{\sigma B}(\sigma)$, as is explained later. In the RMF approximation, the meson field are replaced by the constant mean-field values: $\bar{\sigma}$, $\bar{\omega}$ and $\bar{\rho}$ (the $\rho^0$ field). Note that the mean-field value of the pion vanishes.

In the CQMC model, we assume that the baryon structure variation in matter is reflected in the $\sigma$-field dependence of $g_{\sigma B}(\bar{\sigma})$ \citep[][]{Guichon:1987jp,Saito:1994ki}. For simplicity, we parameterize it in the linear form \citep[][]{Tsushima:1997cu}:
\begin{equation}
	g_{\sigma B}(\bar{\sigma}) = g_{\sigma B}b_{B}\left[1-\frac{a_{B}}{2}(g_{\sigma N}\bar{\sigma})\right],
	\label{eq:sigma-coupling-const}
\end{equation}
where $g_{\sigma B}$ is the $\sigma$-$B$ coupling constant at zero density and $a_B$ and $b_B$ are parameters listed in Table~\ref{tab:parametrizationQMC} \citep[][]{Miyatsu:2011bc}. The parameter $b_B$ represents the effect of the quark-quark hyperfine interaction due to the exchanges of gluon and pion. In the case of $a_{B}=0$ and $b_{B}=1$, the CQMC model is identical to the ordinary RMF model \citep[][]{Serot:1984ey,Bouyssy:1987sh}. Note that the several non-linear self-interaction terms of the scalar ($\sigma$) and/or vector ($\omega, \bm{\rho}, \phi$) mesons \citep[][]{Sugahara:1993wz,Lalazissis:1996rd,ToddRutel:2005zz,Fattoyev:2010mx} are included to soften the EOS and reproduce experimental value for the incompressibility. In contrast, there is no need to introduce the additional meson terms using the CQMC model. Therefore the EOS is determined by $g_{\sigma B}$, $g_{\omega B}$, $g_{\rho B}$, $f_{\omega B}$, $f_{\rho B}$ and $f_{\pi B}$. We investigate the setting for the coupling constants and its validity in Section~\ref{cplcnst}.

In the following, we derive the total energy density of hadronic matter in the RHF approximation. In order to sum up all orders of the tadpole (Hartree) and exchange (Fock) diagrams in the baryon Green's function, $G_B$, we use the Dyson's equation
\begin{equation}
	G_{B}(k) = G_{B}^{0}(k) + G_{B}^{0}(k)\Sigma_{B}(k)G_{B}(k), 
	\label{eq:Dyson-equation}
\end{equation}
with the four momentum of baryon $k^\mu$, the baryon self-energy $\Sigma_{B}$ and the Green's function in free space $G_{B}^{0}$. The baryon self-energy in matter is generally given by
\begin{equation}
	\Sigma_{B}(k) = \Sigma_{B}^{s}(k) - \gamma_{0}\Sigma_{B}^{0}(k)
	+ (\bm{\gamma}\cdot\hat{k})\Sigma_{B}^{v}(k),
	\label{eq:baryon-self-engy}
\end{equation}
where $\hat{k}$ is the unit vector along the (three) momentum $\bm{k}$ and $\Sigma_{B}^{s}$, $\Sigma_{B}^{0}$ and $\Sigma_{B}^{v}$ are the scalar part, the time component of the vector part and the space component of the vector part of the self-energy, respectively. Here we define the effective baryon mass, momentum and energy in matter as \citep[][]{Serot:1984ey,Bouyssy:1987sh}
\begin{eqnarray}
	M_{B}^{\ast}(k) &=& M_{B} + \Sigma_{B}^{s}(k),
	\label{eq:auxiliary-quantity-mass} \\
	k_{B}^{\ast\mu} &=& \left( k_{B}^{\ast0},\bm{k}_{B}^{\ast} \right) 
	= \left(k^{0}+\Sigma_{B}^{0}(k),\bm{k}+\hat{k}\Sigma_{B}^{v}(k) \right),
	\label{eq:auxiliary-quantity-momentum} \\
	E_{B}^{\ast}(k) &=& \left[\bm{k}_{B}^{\ast2}+M_{B}^{\ast2}(k)\right]^{1/2}.
	\label{eq:auxiliary-quantity-engy}
\end{eqnarray}
The baryon self-energies in Equation~(\ref{eq:baryon-self-engy}) are calculated introducing a form factor and their expressions can be seen in \citet{Katayama:2012ge}. Using them, the total baryon number density $n$ and the total energy density of hadronic matter $\varepsilon_H$ are expressed as 
\begin{eqnarray}
	n &=& \sum_{B} \frac{k^{3}_{F_{B}}}{3\pi^{2}},
	\label{eq:baryon-nmb-density} \\
	\varepsilon_{H} &=& \sum_{B}\frac{1}{\pi^{2}}\int_{0}^{k_{F_{B}}} k^2dk~\left[
	T_{B}(k)+\frac{1}{2}V_{B}(k)\right] ,
	\label{eq:baryon-engy-density}
\end{eqnarray}
with
\begin{eqnarray}
	T_{B}(k) &=& \frac{M_{B}M_{B}^{\ast}(k)+kk_{B}^{\ast}}{E_{B}^{\ast}(k)} ,
	\label{eq:baryon-engy-density-kinetic} \\
	V_{B}(k) &=& \frac{M_{B}^{\ast}(k)\Sigma_{B}^{s}(k) 
	+ k_{B}^{\ast}\Sigma_{B}^{v}(k)}{E_{B}^{\ast}(k)}-\Sigma_{B}^{0}(k) ,
	\label{eq:baryon-engy-density-kinetic-potential}
\end{eqnarray}
where $k_{F_{B}}$ is the Fermi momentum of baryon $B$.

\section{Coupling constants}\label{cplcnst}

In this section, we study the coupling constants shown in the previous section. First of all, among the coupling constants, what are not related with hyperons are determined so as to reproduce the gross feature of nuclear mass data with Thomas-Fermi calculation. Then we examine the saturation properties of the resultant EOS. As a next step, we study the coupling constants related with hyperons from the recent analysis of hypernuclei and hyperon production reactions.

\subsection{Thomas-Fermi calculations for atomic nuclei}\label{tf}

In the RHF approximation, the EOS without hyperons is determined by $g_{\sigma N}$, $g_{\omega N}$, $g_{\rho N}$, $f_{\omega N}$, $f_{\rho N}$ and $f_{\pi N}$. We set them to reproduce the experimental data on atomic nuclei. The saturation properties of symmetric nuclear matter are mainly described by the $\sigma$-$N$ and $\omega$-$N$ coupling constants, $g_{\sigma N}$ and $g_{\omega N}$. The $\rho$-$N$ coupling constant $g_{\rho N}$ is related with the symmetry energy. Instead optimizing all the parameters, as for the isoscalar- and isovector-tensor coupling constants, we fix the fractions of $f_{\omega N}$ to $g_{\omega N}$ and $f_{\rho N}$ to $g_{\rho N}$, for simplicity. According to the Nijmegen extended-soft-core (ESC) model \citep[][]{Rijken:2010zzb} based on the $NN$- and $YN$-scattering data, the fractions are $f_{\omega N}/g_{\omega N}=-0.8070/3.5452=-0.2276$ and $f_{\rho N}/g_{\rho N}=3.9298/0.6918=5.6805$. As for $f_{\pi N}$, we use the value suggested by the ESC model. Hereafter, we adopt these values and $g_{\sigma N}$, $g_{\omega N}$ and $g_{\rho N}$ are determined. Note that, in the case of the RHF calculation, not only the symmetry energy but also the saturation energy are sensible to $f_{\rho N}$ \citep[][]{Bouyssy:1987sh}. In contrast, a contribution of $f_{\omega N}$ is tiny near the saturation density. 

In the optimization, the nuclear masses evaluated from our EOS are compared with the experimental mass data. Within the framework of a simplified version of the extended Thomas-Fermi theory \citep[][]{Oyamatsu:2002mv}, the binding energy $B(Z,N)$ of a nucleus with the proton number $Z$ and neutron number $N$ is given by
\begin{equation}
-B(Z,N) = \int \varepsilon_H \left( n_n(\bm{r}), n_p(\bm{r}) \right) d^3 \bm{r} + F_0 \int \left| \nabla n(\bm{r}) \right|^2 d^3 \bm{r} + \frac{e^2}{2} \iint \frac{n_p(\bm{r})n_p(\bm{r}^\prime)}{\left| \bm{r} -\bm{r}^\prime \right|} d^3 \bm{r} d^3 \bm{r}^\prime,
\label{tfbind}
\end{equation}
where $n_n(\bm{r})$ and $n_p(\bm{r})$ are neutron and proton number densities, respectively, and they satisfy their number conservation laws:
\begin{subequations}
\begin{equation}
N = \int n_n(\bm{r}) d^3 \bm{r},
\label{ncon}
\end{equation}
\begin{equation}
Z = \int n_p(\bm{r}) d^3 \bm{r},
\label{pcon}
\end{equation}
\end{subequations}
and $n(\bm{r})$ is a total nucleon number density defined as $n(\bm{r})=n_p(\bm{r})+n_n(\bm{r})$. The first term of the right-hand side of Equation~(\ref{tfbind}) is a bulk term and $\varepsilon_H$ is an energy density of the uniform nuclear matter, which is given by the Hartree-Fock calculation as a function of $n_n$ and $n_p$. The second term corresponds to the gradient energy due to the density inhomogeneity and $F_0$ is the gradient coefficient. While one can treat $F_0$ as an adjustable parameter, we here choose $F_0=68$~MeV~fm$^5$ from previous studies \citep[][]{Oyamatsu:2002mv,Kanzawa:2009uc}. It is confirmed that the result shown below is not sensitive for this choice very much. The third term is Coulomb energy and $e$ is the elementary electric charge. The goal is to determine the density distributions $n_n(\bm{r})$ and $n_p(\bm{r})$ which maximize $B(Z,N)$ under the constraints (\ref{ncon}) and (\ref{pcon}) for given $N$ and $Z$.

For simplicity, we assume spherical nuclei and the nucleon distributions $n_i(r)$ ($i=n,p$) to be \citep[][]{Oyamatsu:2002mv}
\begin{equation}
n_i(r) = \left\{
\begin{array}{ll}
n_i^\mathrm{in} \left[ 1- \left( \frac{r}{R_i} \right)^{t_i} \right]^3, & r < R_i, \\
0, & r \ge R_i,
\end{array}
\right.
\label{profin}
\end{equation}
where $r$ is the distance from the center of the nucleus. In this expression, $R_i$, $t_i$ and $n_i^\mathrm{in}$ ($i=n,p$) are adjustable parameters. Here, $R_i$ roughly represents the nucleon radius, $t_i$ corresponds to the relative surface diffuseness, and $n_i^\mathrm{in}$ is the central nucleon number density. These six parameters are computed for each give nucleus, e.g., $^{208}$Pb ($Z=82$, $N=126$), by the maximization of $B(Z,N)$.

\begin{figure}
\plotone{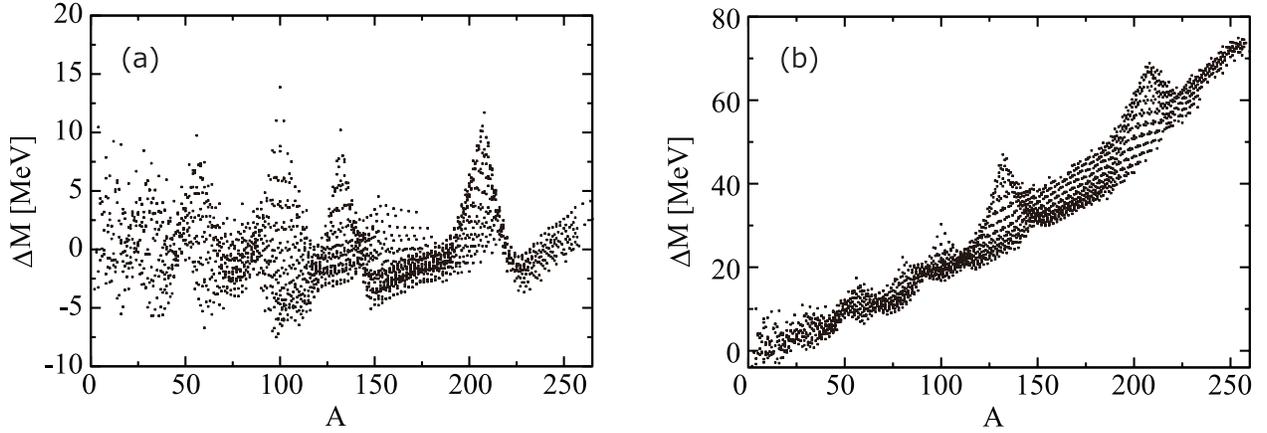}
\epsscale{1.0}
\caption{Mass deviation $\Delta M=M^\mathrm{TF}-M^\mathrm{exp}$ of 2226 nuclei \citep[][]{Audi:2002rp}, where $M^\mathrm{TF}$ is the mass by Thomas-Fermi calculation and $M^\mathrm{exp}$ is the experimental data, for (a) this study and (b) KMS12. The horizontal axes represent the mass number $A$.}
\label{massdv}
\end{figure}

Here, we determine $g_{\sigma N}$, $g_{\omega N}$ and $g_{\rho N}$ so that the binding energies evaluated for 2226 nuclei in the above Thomas-Fermi calculations reproduce the gross feature of their nuclear mass data \citep[][]{Audi:2002rp}. Moreover, in this study, we set the saturation energy to $w_0=-16.1$~MeV according to the previous Thomas-Fermi studies with nuclear data of not only masses but also root-mean-square charge radii \citep[][]{Oyamatsu:2002mv,Kanzawa:2009uc}. Therefore, we regard the combination of $g_{\sigma N}$, $g_{\omega N}$ and $g_{\rho N}$ which satisfies $w_0=-16.1$~MeV and minimizes root-mean-square deviation of the calculated masses from the experimental data as the optimal set. As a result, we find that the minimum value for the root-mean-square deviation is 2.93~MeV, which is comparable with previous studies \citep[][]{Oyamatsu:2002mv,Kanzawa:2009uc}, and the resultant coupling constants are shown in Table~\ref{eosdata} with those of KMS12. In Figure~\ref{massdv}, the differences between the masses by Thomas-Fermi calculations and the experimental data are plotted. We can see that the Thomas-Fermi calculations with the optimal coupling constants in this study well reproduce the gross feature of nuclear mass except shell effects. On the other hand, the root-mean-square deviation given by the Thomas-Fermi calculations with the coupling constants in KMS12 is 36.33~MeV and the mass difference gets roughly larger as the mass number $A$ increases.

\begin{figure}
\epsscale{0.6}
\plotone{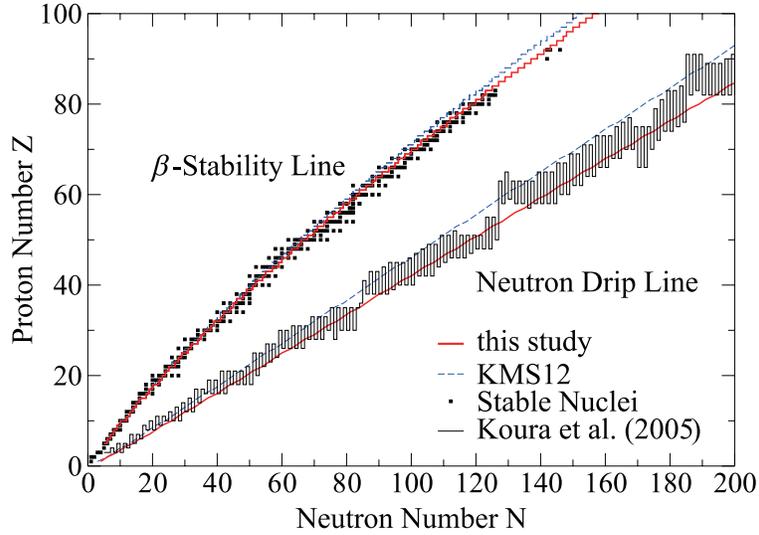}
\caption{$\beta$-stability lines and neutron drip lines obtained from the Thomas-Fermi calculations with the EOS of this study (thick solid) and KMS12 (thick dashed). The regions filled with squares correspond to empirically known stable nuclei \citep[][]{Nuclides:2010}. Thin solid line represents the neutron drip line (with even-odd staggering) from a contemporary mass formula \citep[][]{Koura:2005PTP}.}
\label{dripl}
\end{figure}

The $\beta$-stability line and neutron drip line obtained from the Thomas-Fermi calculations with the EOS of this study are shown in Figure~\ref{dripl}. We can see that the resultant $\beta$-stability line well traces empirically known stable nuclei \citep[][]{Nuclides:2010}. As for the neutron drip line, it is confirmed that our result is consistent with the sophisticated atomic mass formula constructed by \citet{Koura:2005PTP}. We also show the results for the EOS of KMS12. Recently, \citet{Oyamatsu:2010sk} pointed out that the neutron drip line depends on the slope of the symmetry energy, $L$, using EOSs whose Thomas-Fermi calculations reproduce the gross feature of nuclear mass data. In our case, the neutron drip lines of this study and KMS12 differ from each other while, as shown later, the difference of their $L$ is not so large. The neutron drip line for KMS12 may not follow the trend of $L$ because KMS12 is not consistent with mass data. Therefore, the calibration by mass data is important.

\subsection{Saturation properties}\label{satuprop}

Some of the quantities shown in Table~\ref{eosdata} are key parameters for characterizing the saturation properties of uniform bulk nuclear matter. They correspond to the coefficients of the power-series expansions of the energy per baryon $\varepsilon_H (n_n, n_p)/n$ ($n=n_p+n_n$). The energy per baryon of symmetric nuclear matter is written as
\begin{equation}
\frac{\varepsilon_H (n/2, n/2)}{n} = w_0 + \frac{K_0}{18n_0^2} (n-n_0)^2 + {\cal O}(((n-n_0)/n_0)^3),
\label{symmat}
\end{equation}
and that of neutron matter is written as
\begin{equation}
\frac{\varepsilon_H (n, 0)}{n} = \frac{\varepsilon_H (n/2, n/2)}{n} + S_0 + \frac{L}{3n_0} (n-n_0) + {\cal O}(((n-n_0)/n_0)^2).
\label{neumat}
\end{equation}
Both of this study and KMS12 give reasonable values for the saturation density $n_0$ and saturation energy $w_0$. The incompressibility $K_0$ is related to the stiffness of EOS. While the range of $K_0=240 \pm 10$~MeV is led from the isoscalar giant monopole resonance \citep[][]{Piekarewicz:2009gb}, we think that the value of $K_0$ in this study is still reasonable as well as KMS12.

The symmetry energy $S_0$ is the difference between the energies per baryon of symmetric nuclear matter and neutron matter at the saturation density.
Incidentally, some authors define the symmetry energy $E_{\rm sym}(n)$ as 
\begin{equation}
\frac{\varepsilon_H (n_n, n_p)}{n} = \frac{\varepsilon_H (n/2, n/2)}{n} + E_{\rm sym}(n) \delta^2 + {\cal O}(\delta^4),
\label{symdeltaexp}
\end{equation}
where $\delta$ is a neutron-proton asymmetry defined as $\delta \equiv (n_n-n_p)/n$, and use the value at the saturation density, $E_{{\rm sym},0} \equiv E_{\rm sym}(n_0)$. Generally, $E_{{\rm sym},0}$ differs from $S_0$ and both values are shown in Table~\ref{eosdata}.
Recently, the constraint 30~$\mathrm{MeV} \lesssim E_{{\rm sym},0} \lesssim 34$~MeV is suggested by intermediate-energy heavy-ion collisions \citep[][]{Tsang:2008fd} and other works also predict values around $E_{{\rm sym},0} \sim 32$~MeV \citep[][]{baoanli12}. The symmetry energy of this study is consistent with the current implications whereas that of KMS12 is somewhat higher. In Equation~(\ref{neumat}), the parameter $L$ gives the slope of the symmetry energy. It is known that the possible range of $L$ depends on $S_0$ and nuclear models with large values of $S_0$ tend to predict large values of $L$. In particular, a relation $S_0 \approx 28 \, \mathrm{MeV} + 0.075L$ is given by the systematic Thomas-Fermi calculations \citep[][]{Oyamatsu:2002mv}. The value of $L$ in this study satisfies these conditions. According to the recent analyses of terrestrial experiments, the constraints on $L$ favor somewhat smaller value \citep[][]{baoanli12}. However, they have yet to converge and our result is not far from them. Note that, there remain some model dependences in extracting above constraints for the key parameters, and further studies are important.

\subsection{Hyperon potentials in symmetric nuclear matter}\label{hypot}

We determine the coupling constants of hyperons in this subsection. The couplings with $\sigma$ meson are settled from the potential depths of hyperons in symmetric nuclear matter at the saturation density $n_0$ \citep[][]{sb96,Ishizuka:2008gr}. According to experimental results on the single particle energies of many $\Lambda$ hypernuclei, the potential depth of $\Lambda$ in nucleons ($N$) is well estimated as $U^{(N)}_{\Lambda}\simeq-30$~MeV. On the other hand, $\Sigma$ hyperons are thought to feel repulsive potential in nuclear matter as $U^{(N)}_{\Sigma}\sim +30$~MeV from the recently observed quasi-free $\Sigma$ production spectra \citep[][]{harada05,harada06}. It is suggested that a potential depth of $\Xi$ is around $U^{(N)}_{\Xi}\sim -15$~MeV by the analyses of the twin hypernuclear formation \citep[][]{aoki95} and the $\Xi$ production spectra \citep[][]{khaus00}. 

In this study, we set $(U^{(N)}_\Lambda,U^{(N)}_\Sigma,U^{(N)}_\Xi) = (-30$~MeV, $+30$~MeV, $-15$~MeV).
The potential depth of hyperons in nuclear matter is written as \citep[][]{Jaminon:1981xg}
\begin{equation}
	U^{(N)}_{Y} = -g_{\sigma Y} (\bar{\sigma}^{(N)}) \bar{\sigma}^{(N)} + g_{\omega Y}\bar{\omega}^{(N)}
	+\frac{1}{2M_{Y}}\left[-g_{\sigma Y} (\bar{\sigma}^{(N)}) \bar{\sigma}^{(N)} + g_{\omega Y}\bar{\omega}^{(N)} \right]^{2},
	\label{eq:hyperon-potential}
\end{equation}
where $\Lambda$, $\Sigma$ and $\Xi$ are denoted collectively as $Y$. Meanwhile, $\bar{\sigma}^{(N)}$ and $\bar{\omega}^{(N)}$ are the mean-field values of $\sigma$ and $\omega$ mesons in the symmetric nuclear matter, respectively. So as to fit the potential depth for each hyperon in symmetric nuclear matter at $n_{0}$, we determine $g_{\sigma \Lambda}$, $g_{\sigma \Sigma}$ and $g_{\sigma \Xi}$. For this, as in the ESC model \citep[][]{Rijken:2010zzb}, the $\rho$-$Y$ couplings $g_{\rho Y}$ are determined through the SU(6) spin-flavor relations:
\begin{equation}
	g_{\rho N} = \frac{1}{2}g_{\rho \Sigma} = g_{\rho \Xi}, \quad g_{\rho \Lambda} = 0,
	\label{eq:su6}
\end{equation}
where the $\rho$-$N$ coupling $g_{\rho N}$ is fixed in Sec.~\ref{tf}. The tensor couplings of isovector meson $f_{\rho Y}$ are settled again from the fractions of $f_{\rho Y}$ to $g_{\rho Y}$ ($f_{\rho Y}/g_{\rho Y}$) in the ESC model. For the other coupling constants ($g_{\omega Y}$, $f_{\omega Y}$ and $f_{\pi Y}$), we use values in the ESC model for our reference model. Nevertheless, there may be an ambiguity of the $\omega$-$Y$ coupling constants, and we investigate with other values of $g_{\omega Y}$ and $f_{\omega Y}$. The resultant coupling constants of the reference model are summarized in Table~\ref{tab:CouplConst}.

\section{Application for neutron stars}\label{nstar}

In this section, we apply our model to neutron stars. The inside of neutron stars is divided into two parts, the crust and core. In the crust, nucleons distribute nonuniformly while matter is uniform in the core. Firstly we study the neutron star crust with the model described in the former sections to determine the nucleon distributions. Next, we investigate the composition and structure of neutron stars. Here we discuss the impacts of not only the core EOS but also the crust EOS. 

\subsection{Neutron star crust}\label{crust}

In the neutron star crust, the lattice of proton clusters is structured and the bcc lattice is preferred to minimize the Coulomb energy \citep[][]{Oyamatsu:1984PTP}. Neutrons accompany the proton distribution and form ``nuclei''. In high density regime, a part of neutrons drip out of the ``nuclei''. The region where the neutron drip occurs is called inner crust. The low density region without the neutron drip is called outer crust. Electrons exist so as to achieve the charge neutrality and $\beta$-equilibrium, and distribute almost uniformly due to their high Fermi energy.

In this paper, we study the nucleon distribution in the crust extending Thomas-Fermi model described in Sec.~\ref{tf} \citep[][]{Oyamatsu:1993zz,Shen:1998gq,Kanzawa:2009uc}. We treat the bcc lattice approximately with a spherical Wigner-Seitz cell. Each unit cell has the same volume $a^3$ and we refer to $a$ as the lattice constant. The total energy in the cell is written as 
\begin{equation}
W = W_N + W_e + W_\mathrm{Coul}.
\label{tfcrust}
\end{equation}
The first term $W_N$ is the nuclear energy and expressed, as in Equation~(\ref{tfbind}), in the density functional form:
\begin{equation}
W_N = \int_\mathrm{cell} \left\{ \varepsilon_H \left( n_n(\bm{r}), n_p(\bm{r}) \right) + M_n n_n(\bm{r}) + M_p n_p(\bm{r}) + F_0 \left| \nabla n(\bm{r}) \right|^2 \right\} d^3 \bm{r},
\label{tfcrustn}
\end{equation}
where $M_n$ and $M_p$ are the neutron mass and proton mass, respectively. Note that the terms, $M_n n_n(\bm{r})$ and $M_p n_n(\bm{r})$, correspond to the rest mass energy of nucleons. The second term $W_e$ is the electron energy. In this paper, we regard electrons as uniform relativistic Fermi gas and $W_e$ is calculated as
\begin{equation}
W_e = \frac{m_e^4a^3}{8\pi^2} \left\{ x_e (2x_e^2+1)(x_e^2+1)^{1/2} - \ln \left[ x_e+(x_e^2+1)^{1/2} \right] \right\},
\label{tfcruste}
\end{equation}
with
\begin{equation}
x_e = \frac{1}{m_e} \left( 3\pi^2n_e \right)^{1/3},
\label{xe}
\end{equation}
where $m_e$ and $n_e$ are the electron mass and electron number density. Note that $n_e$ is determined so as to satisfy the charge neutrality condition,
\begin{equation}
a^3 n_e = Z = \int_\mathrm{cell} n_p(\bm{r}) d^3 \bm{r}.
\label{chargen}
\end{equation}
with the proton number in the cell, $Z$. The last term $W_\mathrm{Coul}$ is the Coulomb energy in the Wigner-Seitz cell and written as
\begin{equation}
W_\mathrm{Coul} = \frac{e^2}{2} \iint \frac{[n_p(\bm{r})-n_e][n_p(\bm{r}^\prime)-n_e]}{\left| \bm{r} -\bm{r}^\prime \right|} d^3 \bm{r} d^3 \bm{r}^\prime + c_\mathrm{bcc}\frac{(Ze)^2}{a},
\label{tfcrustc}
\end{equation}
where the second term on the right-hand side of Equation~(\ref{tfcrustc}) is the correction for the bcc lattice with $c_\mathrm{bcc}=0.006562$ \citep[][]{Oyamatsu:1993zz}.

For nucleon distributions in the Wigner-Seitz cell, we assume spherical symmetry and again utilize the parameterization:
\begin{equation}
n_i(r) = \left\{
\begin{array}{ll}
(n_i^\mathrm{in}-n_i^\mathrm{out}) \left[ 1- \left( \frac{r}{R_i} \right)^{t_i} \right]^3 + n_i^\mathrm{out}, & r < R_i, \\
n_i^\mathrm{out}, & R_i \le r \le R_\mathrm{cell},
\end{array}
\right.
\label{profcrust}
\end{equation}
where $R_i$, $t_i$, $n_i^\mathrm{in}$ and $n_i^\mathrm{out}$ ($i=n,p$) are adjustable parameters. The radius of Wigner-Seitz cell $R_\mathrm{cell}$ is related to the lattice constant $a$ as: $R_\mathrm{cell}=(3/4\pi)^{1/3}a$. Note that $n_n^\mathrm{out}$ corresponds to the number density of dripped neutrons and we set $n_p^\mathrm{out}=0$. Here the ground state of the system at given (average) baryon number density,
\begin{equation}
n_B = \frac{1}{a^3}\int_\mathrm{cell} n(\bm{r}) d^3 \bm{r},
\label{nbdef}
\end{equation}
is determined minimizing the total energy density $\varepsilon=W/a^3$. When the minimized total energy density is higher than that of uniform matter at the same $n_B$, we regard the uniform matter as the ground state.

\begin{figure}
\epsscale{0.5}
\plotone{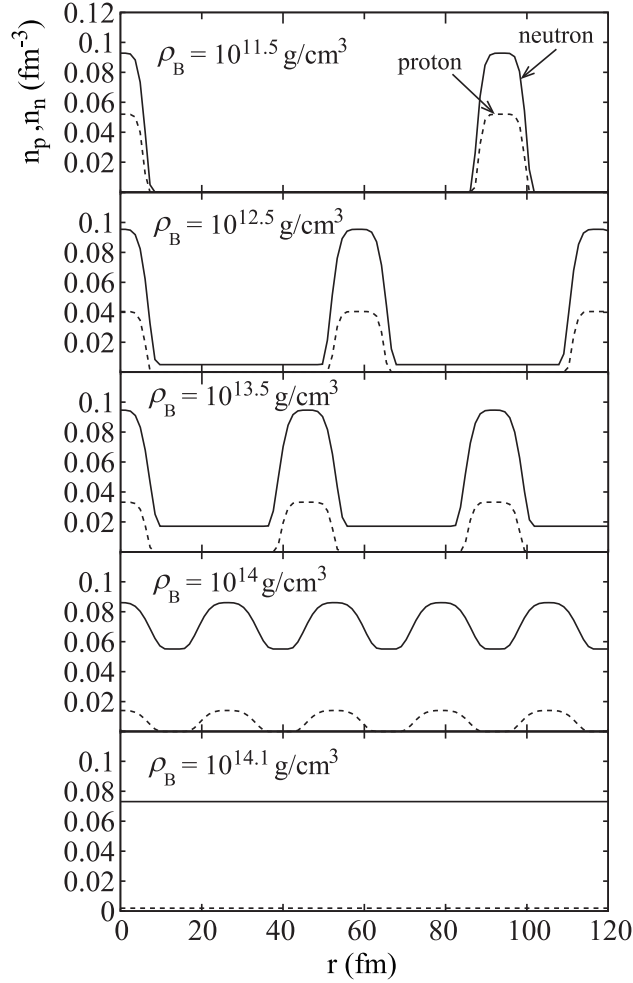}
\caption{The neutron distribution (solid) and proton distribution (dashed) along the straight line joining the centers of the nearest nuclei in the bcc lattice. The plots correspond, from top to bottom, to the cases at the baryon mass density $\rho_B=10^{11.5}$, $10^{12.5}$, $10^{13.5}$, $10^{14.0}$ and $10^{14.1}$~g~cm$^{-3}$ ($n_B=1.90\times10^{-4}$, $1.90\times10^{-3}$, $1.90\times10^{-2}$, $6.02\times10^{-2}$, $7.58\times10^{-2}$~fm$^{-3}$ in baryon number density).}
\label{npdist}
\end{figure}

We show the resultant nucleon distributions in Figure~\ref{npdist}. The distance between nearest nuclei becomes smaller with increase of the density. Above the critical density $n_\mathrm{drip}$, the neutron drip occurs. For our model, it is evaluated as $n_\mathrm{drip}=2.70\times10^{-4}$~fm$^{-3}$, which corresponds to $4.48\times10^{11}$~g~cm$^{-3}$ in baryon mass density. At a higher density, nuclei melt and the phase transition from nonuniform matter to uniform matter takes place. The transition density is $n_\mathrm{uni}=6.85\times10^{-2}$~fm$^{-3}$ ($1.14\times10^{14}$~g~cm$^{-3}$ in baryon mass density) for our model. Note that, for a region with somewhat lower density than $n_\mathrm{uni}$, nuclei are thought to deform to rodlike and slablike shapes, which are often called nuclear pasta \citep[][]{Ravenhall:1983uh,Hashimoto:1984PTP}. The pasta nuclei may further reduce the energy density. While we do not take into account pasta nuclei, they would not give significant change to the EOS of matter. Moreover, since the slope of the symmetry energy is somewhat high as $L=77.1$~MeV for our model, the density region containing pasta nuclei would not so large \citep[][]{Oyamatsu:2006vd}.

\begin{figure}
\epsscale{0.45}
\plotone{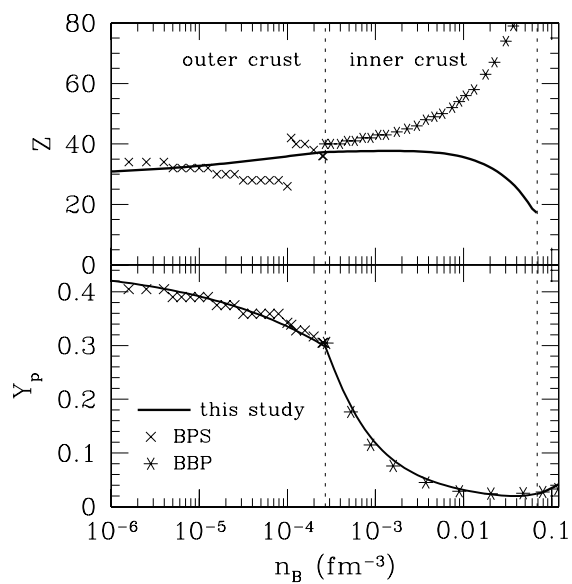}
\caption{Proton number of nuclei (top panel) and average proton fraction (bottom panel) in the neutron star crust as functions of baryon number density. In both panels, the results of our Thomas-Fermi calculations in this study (solid lines) are compared with the models of BPS and BBP (plots). Vertical dotted lines show the boundaries between outer crust and inner crust and between inner crust and core.}
\label{crustzx}
\end{figure}

In Figure~\ref{crustzx}, we show the results of our Thomas-Fermi calculations for proton number $Z$ determined by Equation~(\ref{chargen}) and the average proton fraction $Y_p$ given by
\begin{equation}
Y_p = \frac{\int_\mathrm{cell} n_p(\bm{r}) d^3 \bm{r}}{\int_\mathrm{cell} n(\bm{r}) d^3 \bm{r}}.
\label{ypdef}
\end{equation}
We compare them with the model of \citet{Baym:1971pw}, hereafter BPS, for the outer crust region and the model of \citet{Baym:1971ax}, hereafter BBP, for the inner crust region. The calculation of BPS is based on a mass formula taking into account the even-odd term and shell effect, which are not dealt in our model. Therefore, the preference of the magic nuclei can be seen for BPS (for instance, the jump of $Z$ near $n_B \sim 10^{-4}$~fm$^{-3}$ in Figure~\ref{crustzx}) while $Z$ is smooth as a function of $n_B$ for our model. However, the results for $Y_p$ are similar between BPS and this study. On the other hand, the proton number of nuclei in the inner crust of this study is quite different from that of BBP, in which a compressible liquid-drop model is used. According to the systematic study by \citet{Oyamatsu:2006vd}, for the inner crust region, the value of $Z$ strongly depends on the slope of the symmetry energy, $L$. The discrepancy between BBP and our model would be within this uncertainty. Incidentally, the frequency of quasi-periodic oscillations discovered in decaying tail of the giant flares, which are bursty gamma-ray emission from neutron stars with strong magnetic fields, is sensitive to $Z$. Recently, \citet{Sotani:2012qc} pointed out that low $Z$ models, or EOSs with $L\gtrsim50$~MeV, would be preferred to account for the observed frequency from SGR 1806-20. Note that, the results for $Y_p$ are again similar between BBP and this study.

\subsection{Composition and structure of neutron stars}\label{core}

\begin{figure}
\begin{center}
\includegraphics[scale=0.5,angle=270]{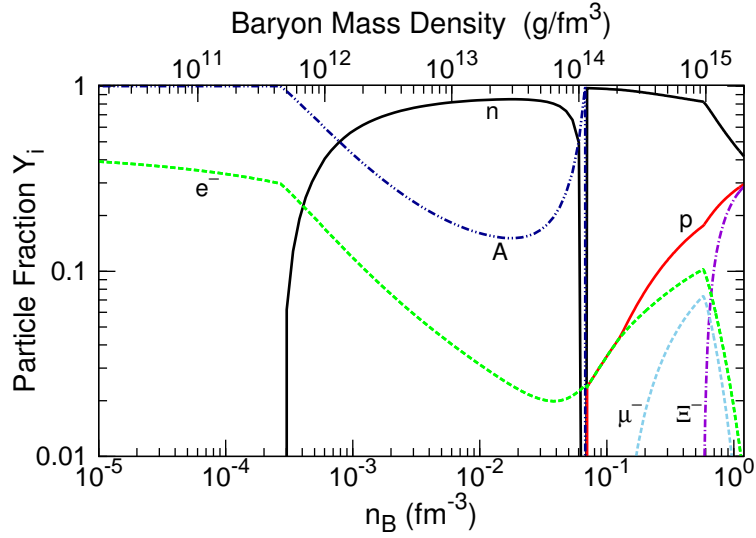}
\caption{Particle fractions, $Y_{i}$, of the neutron star matter as functions of the total baryon number density, $n_{B}$, from the crust region to the core region. They are defined as $Y_{i} = n_{i}/n_{B}$ with the number densities of particle $i$, $n_{i}$, except for the nucleus fraction, which is defined by $Y_A=A/(n_{B} a^3)$ with the volume of Wigner-Seitz cell $a^3$ and the mass number of nucleous $A$ as in \citet{Shen:2011qu}.}
\label{fig:composition}
\end{center}
\end{figure}

Nuclei melt into uniform matter of neutrons and protons in the neutron star core. In this region, not only electrons but also muons reside achieving the charge neutrality and $\beta$-equilibrium under weak processes. Here we treat them as non-interacting fermions. Furthermore, hyperons are also generated at higher densities. In Figure~\ref{fig:composition}, we show the particle fractions, $Y_{i}=n_{i}/n_{B}$, from the crust region to the core region as functions of the total baryon number density, $n_{B}$, where $n_i$ represents the number density of particle $i$. As already stated, there are dripped neutrons in the inner crust. Note that, the fraction of nuclei ($A$) increases artificially near the transition density to the uniform matter, $n_\mathrm{uni}$, because the radius of nuclei, $R_n$, gets closer to that of Wigner-Seitz cell, $R_\mathrm{cell}$, in Equation~(\ref{profcrust}). Not the spherical nuclei assumed here but the pasta nuclei may reside in this region.

\begin{figure}
\begin{center}
\includegraphics[scale=0.5,angle=270]{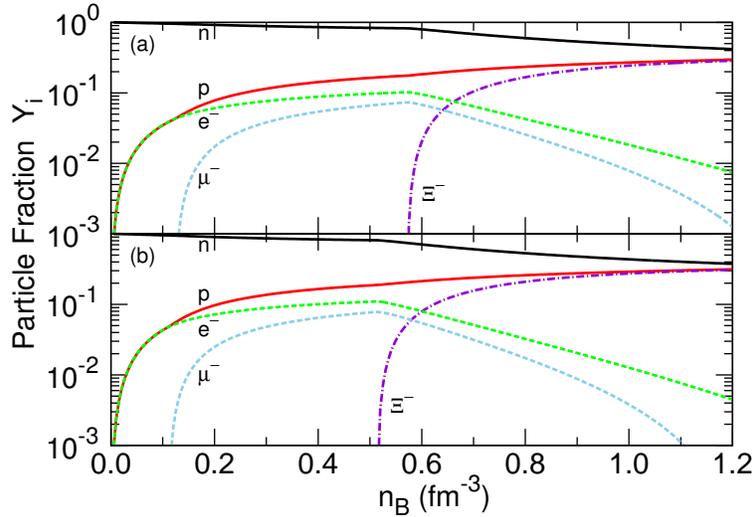}
\caption{Particle fractions of the uniform neutron star matter for the models of (a) this study and (b) KMS12.}
\label{fig:composition-core}
\end{center}
\end{figure}

We show the particle fractions of the uniform neutron star matter in Figure~\ref{fig:composition-core}. In the RHF calculation of this study, since the Fock contribution enhances the repulsive effect mainly through $\omega$ mesons at supra-nuclear densities, the hyperon creation is suppressed. Therefore the threshold densities of hyperons rise. In contrast, the creation of $\Xi^{-}$ is promoted by the inclusion of the tensor coupling with $\rho$ mesons \citep[][]{RikovskaStone:2006ta,whittenbury12}. In fact, the critical density of $\Xi^{-}$ creation in this study is higher than KMS12. Since the value of $f_{\rho \Xi}/g_{\rho \Xi}$ is fixed and the coupling relation in Equation~(\ref{eq:su6}) is assumed both in our model and KMS12, the value of $f_{\rho \Xi}/g_{\rho N}$ is kept. Therefore our model has a smaller absolute value of $f_{\rho \Xi}$ comparing with KMS12 (see also Table~\ref{eosdata}), which is consistent with the difference in the critical density between two models. Note that $\Lambda$ hyperons do not interact with $\rho$ mesons due to $\bm{I}_\Lambda=0$ in Equation~(\ref{eq:Lagrangian-interaction}). In addition, we assume that the $\Sigma$ hyperons feel a repulsive interaction at high densities and, hence, they cannot appear. As a result, among the hyperons taken into account, only $\Xi^{-}$ appears and the others are not produced below 1.2~fm$^{-3}$. 

\begin{figure}
\begin{center}
\includegraphics[scale=0.5,angle=270]{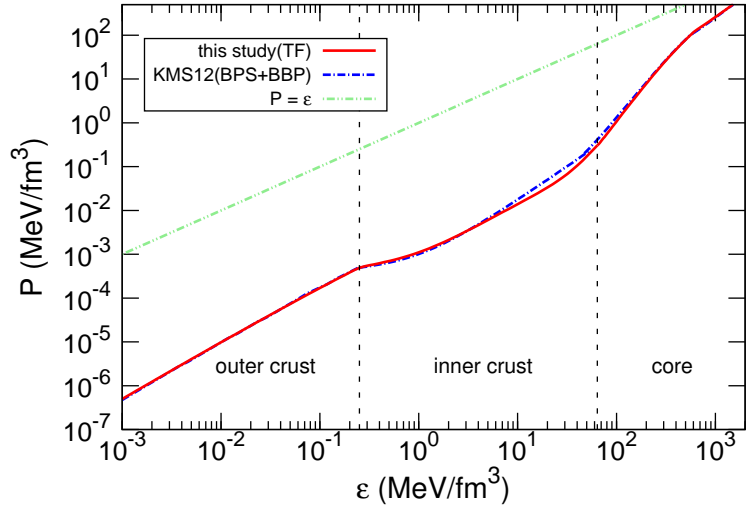}
\caption{Equation of state (pressure vs. energy density) in the crust and core regions. Solid and dot-dashed lines correspond to EOSs in this study and KMS12, respectively. The meaning of vertical dotted lines is same as Figure~\ref{crustzx}.}
\label{fig:EOS}
\end{center}
\end{figure}

In Figure~\ref{fig:EOS}, we show the EOS for neutron star matter from the crust region to the core region. The pressure is given as
\begin{equation}
P=n_{B}^{2}\frac{\partial}{\partial n_{B}}\left(\frac{\varepsilon}{n_{B}}\right).
\label{eq:pressure}
\end{equation}
In this study, the boundary between the crust and core is determined self-consistently comparing the energy density of nonuniform matter with that of uniform matter. Therefore, our EOS continues smoothly from the crust to the core. In contrast, the previous EOS of KMS12 adopts BPS model \citep[][]{Baym:1971pw} for the outer crust and BBP model \citep[][]{Baym:1971ax} for the inner crust. In the inner crust region, the size of nuclei is quite different between our model and BBP model as shown in Figure~\ref{crustzx}, and it is reflected in EOS. On the other hand, in the core region, our model is stiffer than the EOS of KMS12 while they are barely distinguishable in the figure. This is consistent to the fact that, as shown in Table~\ref{eosdata}, our EOS has large incompressibility $K_0$ than the EOS of KMS12. In the higher density region, we can see the softening of EOS due to the mixture of $\Xi^{-}$ hyperons.
Note that, our EOS satisfies the causality even at high densities because it is based on the relativistic framework. Incidentally EOSs constructed by nonrelativistic theories \citep[e.g.,][]{kann07,Kanzawa:2009uc} sometimes cause the causality violation, which corresponds to the region above the line of $P=\varepsilon$ in Figure~\ref{fig:EOS}.

\begin{figure}
\begin{center}
\includegraphics[scale=0.5,angle=270]{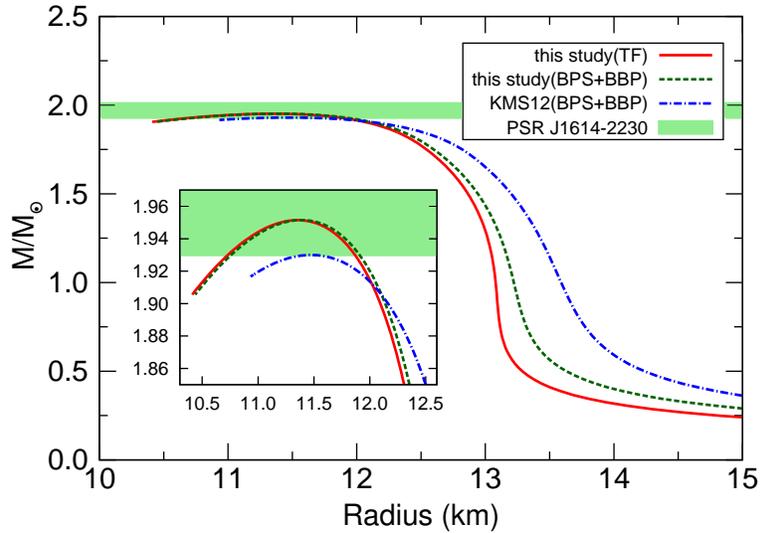}
\caption{Mass-radius relations of neutron stars. Solid line shows the result of this study where the crust EOS is calculated by the Thomas-Fermi model, and dashed line corresponds to the model in which the core EOS is the same as our model and BPS \citep[][]{Baym:1971pw} and BBP \citep[][]{Baym:1971ax} models are adopted as the crust EOS. Dot-dashed line is for KMS12, which adopts BPS and BBP for the crust. The shaded area is the mass range of PSR J1614-2230, $1.97\pm0.04M_{\odot}$ \citep[][]{Demorest:2010bx}.}
\label{fig:TOV}
\end{center}
\end{figure}

In Figure~\ref{fig:TOV}, the neutron star radii $R$ are shown as functions of masses $M$. The result of KMS12 is drawn for comparison, as well as that of the model in which the core EOS is the same as our model and BPS and BBP models are adopted as the crust EOS. It is known that the inclusion of hyperons generally reduces the maximum mass of neutron stars drastically \citep[][]{Glendenning:1991es,SchaffnerBielich:2008kb}. Nevertheless, the Fock contribution prevents the hyperon appearance in our model, and the resultant maximum mass is high as $1.95M_{\odot}$. This value is within the mass range of recently observed massive neutron star J1614-2230, $1.97\pm0.04M_{\odot}$ \citep[][]{Demorest:2010bx}.\footnote{After the first submission of this paper, \citet[][]{Antoniadis:2013pzd} reported that the mass of PSR J0348+0432 was evaluated as $2.01\pm0.04M_{\odot}$ in the $1\sigma$ limit. Their analysis depends on the model of the companion white dwarf and further investigation is important.} Incidentally, the maximum mass of our model is higher than that of KMS12 ($1.93M_{\odot}$) for the two reasons. The first one is because the critical density of $\Xi^{-}$ creation is higher for our model. For the second reason, as already stated, our EOS is stiffer than the EOS of KMS12. In our model, $\Xi^{-}$ hyperons are included in the core of neutron stars more massive than $1.77M_{\odot}$, and the maximum-mass neutron star has the baryon number density of 1.01~fm$^{-3}$ ($1.67\times10^{15}$~g~cm$^{-3}$ in baryon mass density) at the center.

\begin{figure}
\begin{center}
\includegraphics[scale=0.5,angle=270]{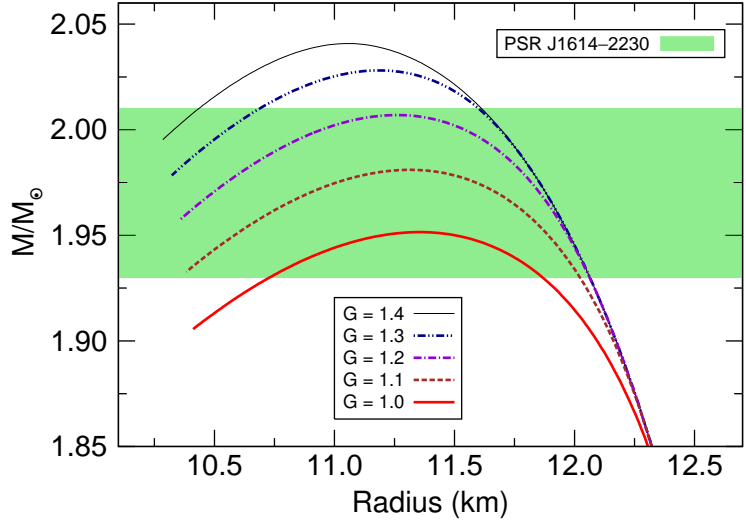}
\caption{Mass-radius relations of neutron stars for the model with various values of the coupling ratio $G=g_{\omega Y}/g_{\omega Y}^{\rm ESC}=f_{\omega Y}/f_{\omega Y}^{\rm ESC}$. The lines correspond, from bottom to top, to the cases with $G=1.0$, $1.1$, $1.2$, $1.3$ and $1.4$. The shaded area is the same as Figure~\ref{fig:TOV}. The crust EOS calculated by the Thomas-Fermi calculates is adopted among all models}
\label{fig:TOVa}
\end{center}
\end{figure}

The mass-radius relation of neutron stars is determined by EOS. We examine the dependence of the maximum mass on the $\omega$-$Y$ coupling constants, as mentioned in Sec.~\ref{hypot}. As for $g_{\omega Y}$ and $f_{\omega Y}$, we adopt the values suggested by the ESC model, $g_{\omega Y}^{\rm ESC}$ and $f_{\omega Y}^{\rm ESC}$, in our reference model. Here we investigate the dependence on the coupling ratio $G$, which is defined as the ratio of couplings to ESC values, $G=g_{\omega Y}/g_{\omega Y}^{\rm ESC}=f_{\omega Y}/f_{\omega Y}^{\rm ESC}$. In Figure~\ref{fig:TOVa}, we show the variation in mass-radius relations of neutron stars with the coupling ratio. As the coupling ratio increases, the creation densities of $\Xi^{-}$ become higher and the maximum mass of neutron stars gets larger. Furthermore, for the cases with $G>1.4$, hyperons do not appear even in the core of maximum-mass neutron star. Therefore, the maximum mass does not become larger than $2.04M_{\odot}$ in our models. Note that, there is no $G$ dependence of the low density EOS including the crust region.

\begin{figure}
\epsscale{0.45}
\plotone{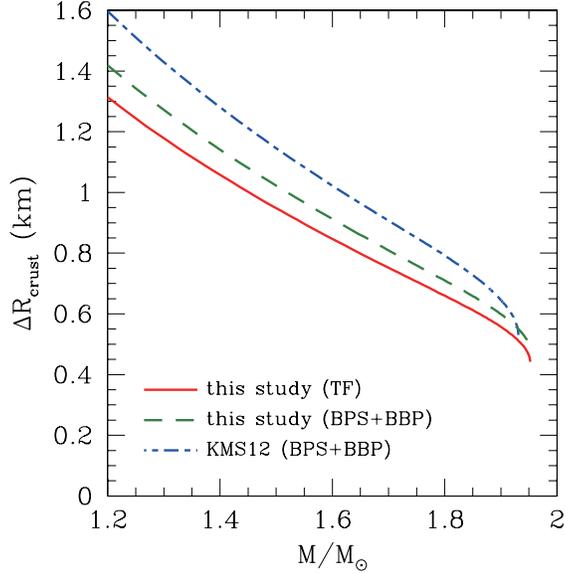}
\caption{Crust thickness as a function of neutron star mass. The notations of lines are the same as Figure~\ref{fig:TOV}.}
\label{crustthickness}
\end{figure}

In our model, the radii of neutron stars are around 12-13~km for typical values of the neutron star mass. The difference between our EOS and KMS12 comes mainly from the core region. On the other hand, as seen in Figure~\ref{fig:TOV}, the crust EOS also affects the radius at about 1\%. When a neutron star radius is observationally determined in high precision, we may be able to study not only the core EOS but also the crust EOS. Furthermore, as for the crust thickness $\Delta R_\mathrm{crust}$, the impact of crust EOS is comparable with that of core EOS. To see this more clearly, we show $\Delta R_\mathrm{crust}$ as functions of neutron star masses in Figure~\ref{crustthickness}. The variation is about 10\% among the different EOS models. Incidentally, the pulsar glitches are thought to associate with the superfluidity of dripped neutrons in the inner crust and the averaged rate of spin-reversal due to glitches depends on the moment of inertia of superfluid neutrons \citep[][]{AI75,Link99,Nils12}. To determine the neutron star structure and composition precisely, the crust EOS is important as well as the core EOS.

\section{Summary}\label{summary}
% \section{Summary and discussion}\label{summary}
We have constructed the EOS for neutron star matter at zero temperature including nuclei in the crust and hyperons in the core. For the EOS of uniform nuclear matter, the framework of the CQMC model based on the RMF theory, which was proposed in \citet{Katayama:2012ge}, has been adopted. In this study, the $\sigma$-$N$, $\omega$-$N$ and $\rho$-$N$ couplings have been determined so as to reproduce the gross feature of nuclear mass data with Thomas-Fermi calculation. The $\sigma$-$Y$ couplings have been determined to fit the potential depth in symmetric nuclear matter from the recent analyses of hypernuclei and hyperon production reactions. The $\omega$-$Y$ and $\pi$-$B$ couplings have been taken from the ESC model based on the hyperon-nucleon scattering data. The $\rho$-$Y$ couplings have been determined through the SU(6) spin-flavor relations. The tensor couplings have been determined from scaling of the ESC model. Due to the calibration from the mass data, the resultant symmetric energy is consistent with the current implications.

To construct the neutron star EOS, we have dealt with not only high-density uniform matter in the core but also low-density nonuniform matter in the crust within the Thomas-Fermi approximation. In our method, the neutron drip and transition to uniform matter are described self-consistently. Although all octet baryons are taken into account in the core region, only $\Xi^{-}$ appears and the other hyperons are not generated below 1.2~fm$^{-3}$. The resultant maximum mass of neutron stars is $1.95M_\odot$, which is consistent with the mass range of recently observed massive neutron star J1614-2230, $1.97\pm0.04M_{\odot}$ \citep[][]{Demorest:2010bx}. As a future work, we will extend this study to finite temperatures for use in astrophysics.

\acknowledgments

The authors express their appreciation to Koichi Saito and Hideyuki Suzuki for continuing encouragements. One of the authors (T.M.) thanks Myung-Ki Cheoun and Chung-Yeol Ryu for fruitful discussions on the EOS for neutron star matter. One of the authors (K.N.) is grateful to Kei Iida and Kazuhiro Oyamatsu for useful discussions and valuable comments. This work was partially supported by Grants-in-Aids for Research Activity Start-up through No.~23840038 provided by the Japan Society for Promotion of Science (JSPS) and for Scientific Research on Innovative Areas through No.~24105008 provided by the Ministry of Education, Culture, Sports, Science and Technology (MEXT) in Japan.

\appendix
\section{EOS Table} \label{eostable}
The results for the reference model in this study are publicly available on the Web at\\
{\tt http://asphwww.ph.noda.tus.ac.jp/myn/}\\
This data set is open for general use in any numerical simulations for astrophysics. There are not only the EOS table but also the data on particle fractions and properties of nuclei in the crust region.

\clearpage

\begin{deluxetable}{lcccc}
\tablewidth{0pt}
\tablecaption{Values of $a_B$ and $b_B$ in the CQMC model.}
\tablehead{ $B$ & $N$ & $\Lambda$ & $\Sigma$ & $\Xi$ }
\startdata
$a_{B}$~(fm)& 0.118 & 0.122 & 0.184 & 0.181 \\
$b_{B}$     & 1.04  & 1.09  & 1.02  & 1.15
\enddata
\label{tab:parametrizationQMC}
\end{deluxetable}

\begin{deluxetable}{lccccccccc}
\tabletypesize{\footnotesize}
\tablewidth{0pt}
\tablecaption{Coupling constants and results for saturation properties.}
\tablehead{ & & & & $w_0$ & $n_0$ & $S_0$ & $E_{{\rm sym},0}$ & $K_0$ & $L$ \\
 EOS & $g_{\sigma N}/\sqrt{4\pi}$ & $g_{\omega N}/\sqrt{4\pi}$ & $g_{\rho N}/\sqrt{4\pi}$ & (MeV) & (fm$^{-3}$) & (MeV) & (MeV) & (MeV) & (MeV) }
\startdata
 this study & 1.94 & 2.39 & 0.596 & $-16.1$ & 0.155 & 33.6 & 32.7 & 274 & 77.1 \\
 KMS12      & 1.81 & 2.42 & 0.692 & $-15.7$ & 0.150 & 36.5 & 35.5 & 261 & 86.0
\enddata
\label{eosdata}
\tablecomments{The value of $L$ for KMS12 is different from 84.8~MeV, the result shown as case (d) in Table~8 of \citet{Katayama:2012ge}, because the definitions are different. We define the slope of the symmetry energy $L$ by Equation~(\ref{neumat}), while KMS12 defined it by the density derivative of $E_{\rm sym}(n)$. With their definition, the model in this study gives $L=75.8$~MeV.}
\end{deluxetable}

\begin{deluxetable}{lcccc}
\tablewidth{0pt}
\tablecaption{Coupling constants used in the reference model of this study.}
\tablehead{ $B$ & $N$ & $\Lambda$ & $\Sigma$ & $\Xi$ }
\startdata
$g_{\sigma B}/\sqrt{4\pi}$ &     1.94 &    2.15 &     1.67 &      1.50 \\
$g_{\omega B}/\sqrt{4\pi}$ &     2.39 &    2.82 &     2.82 &      2.09 \\
$f_{\omega B}/\sqrt{4\pi}$ & $-0.545$ & $-3.39$ & $-0.261$ &   $-4.40$ \\
$g_{\rho B}/\sqrt{4\pi}$   &    0.596 &       0 &     1.19 &     0.596 \\
$f_{\rho B}/\sqrt{4\pi}$   &     3.39 &       0 &     2.94 &  $-0.446$ \\
$f_{\pi B}/\sqrt{4\pi}$    &    0.268 &       0 &    0.190 & $-0.0772$
\enddata
\label{tab:CouplConst}
\end{deluxetable}

\end{document}